\documentclass[a4paper,11pt]{article}
\usepackage{jheppub}
\usepackage{amsmath,amssymb,bm,graphicx,bbold,epsf,colordvi}
\usepackage{lipsum}
\usepackage{soul}
\usepackage{braket}
\usepackage{bm}
\allowdisplaybreaks 
\usepackage{color}
\usepackage[dvipsnames]{xcolor}

\usepackage{mathrsfs}
\usepackage{soul} 
\allowdisplaybreaks
\usepackage[abs]{overpic}
\usepackage[export]{adjustbox}
\usepackage{bm}
\usepackage{scalerel}
\usepackage{accents}

\usepackage{slashed}
\newcommand{\bef}{\begin{figure}[hbt]\centering}
\newcommand{\eef}{\end{figure}}
\usepackage{mathtools}
\usepackage{subfigure}
\usepackage{booktabs}
\usepackage{graphicx}
\graphicspath{ {./figure/} }
\usepackage{comment}
\usepackage{lipsum}

\def\d{\text{d}}

\newcommand{\nf}{{(n_f)}}
\newcommand{\nl}{{(n_l)}}

\newcommand{\cf}{C_F}
\newcommand{\ca}{C_A}

\newcommand{\beq}{\begin{equation}}
\newcommand{\eeq}{\end{equation}}
\def\bea#1\eea{\begin{align}#1\end{align}}

\def \be  {\begin{equation}}
\def \ee  {\end{equation}}
\def \ba  {\begin{eqnarray}}
\def \ea  {\end{eqnarray}}
\newcommand\as{\alpha_s}

\allowdisplaybreaks

\newcommand{\md}{\mathrm{d}}

\newcommand{\mS}{\mathcal{S}}
\newcommand{\mJ}{\mathcal{J}}

\newcommand{\oat}{\mathcal{O}(\alpha_s^2)}
\allowdisplaybreaks[4]

\makeatletter
\def\@fpheader{~}
\makeatother

\usepackage[Q=yes,pverb-linebreak=no]{examplep}

\title{NNLL$^\prime$ resummation of azimuthal decorrelation for boosted top quark pair production at the LHC}

\author[a]{Qian-Shun Dai}
\author[a]{, Ming-Jun Liu}
\author[a,b,c]{, and Ding Yu Shao}

\affiliation[a]{Department of Physics, Center for Field Theory and Particle Physics and Key Laboratory of Nuclear Physics and Ion-beam Application (MOE), Fudan University, Shanghai, 200433, China}
\affiliation[b]{Shanghai Research Center for Theoretical Nuclear Physics, NSFC and Fudan University, Shanghai 200438, China}
\affiliation[c]{Center for High Energy Physics, Peking University, Beijing 100871, China}

\emailAdd{qsdai23@m.fudan.edu.cn, mjliu24@m.fudan.edu.cn, dyshao@fudan.edu.cn}

\abstract{
The precision program of the Large Hadron Collider (LHC) increasingly relies on the boosted regime, where top quark properties are probed at the TeV scale. However, the simultaneous presence of heavy quark mass effects and large logarithmic corrections from soft radiation poses a significant challenge for theoretical predictions. In this work, we develop a transverse momentum dependent (TMD) factorization and resummation framework for boosted top quark pair production in the back-to-back limit at the LHC. By employing a two step matching procedure, matching QCD through $\mathrm{SCET}\,+\,\mathrm{HQET}$ onto $\mathrm{SCET}\,+\,\mathrm{bHQET}$, we systematically resum large logarithms associated with both the top quark mass and the azimuthal decorrelation. A key component of our formalism is the first extraction of the two-loop ultra-collinear function, obtained via the refactorization of the fully differential massive soft function. This result completes the set of perturbative ingredients required to achieve $\mathrm{NNLL}^\prime$ accuracy for the azimuthal decorrelation distribution. Our framework establishes a new benchmark for heavy-quark TMD resummation in the boosted limit at hadron colliders.
}

\begin{document}
\maketitle
\flushbottom

\section{Introduction}
\label{sec:intro}

The top quark, the heaviest elementary particle in the Standard Model (SM), occupies a unique position in high-energy physics. With a mass at the electroweak scale, its large Yukawa coupling makes it an indispensable probe of the electroweak symmetry-breaking sector and the dynamics of quantum chromodynamics (QCD). Because its lifetime is shorter than the characteristic timescale of hadronization, the top quark decays as a quasi-free particle, ensuring that its spin information is preserved and directly transferred to its decay products. At the Large Hadron Collider (LHC), the combination of large production cross sections and a vast kinematic reach enables high precision tests of the SM and searches for new physics across multiple energy scales.

Of particular interest are processes in the boosted regime, where top quarks are produced with transverse momenta or pair invariant masses far exceeding the top-quark mass, $m_t$. This high-energy frontier, extending into the multi-TeV scale, serves as a primary arena for manifesting new massive states predicted by various beyond the SM scenarios. Given the top quark's large mass, many such models predict new heavy resonances that couple preferentially to the third generation, yielding distinct signatures in the large invariant mass spectrum. Furthermore, precision measurements of angular and spin correlation observables in this regime offer sensitivity to subtle deviations from SM predictions, such as those arising from modified top quark couplings or new contact interactions. However, the emergence of disparate physical scales in this regime poses significant challenges for theoretical predictions, necessitating a systematic treatment of large logarithmic corrections.

Boosted top-quark production also offers a promising avenue for the precise determination of the top-quark mass. As a fundamental parameter for electroweak precision tests \cite{ALEPH:2005ab, Awramik:2003rn, Baak:2012kk, Baak:2014ora}, Higgs boson physics \cite{Haller:2018nnx}, and the stability of the electroweak vacuum \cite{Degrassi:2012ry, Buttazzo:2013uya, Branchina:2013jra, Bednyakov:2015sca, Andreassen:2017rzq}, $m_t$ must be measured with the highest possible accuracy. Conventional measurements, however, are often limited by ambiguities in relating experimental observables to a well-defined field theoretic mass scheme \cite{Beneke:1999qg, Hoang:2008xm}. In this context, theoretically clean observables accessible in the boosted regime, such as event shapes \cite{Fleming:2007xt, Fleming:2007qr, Hoang:2008xm} and energy energy correlators \cite{Holguin:2022epo, Holguin:2023bjf, Holguin:2024tkz, Moult:2025nhu}, provide a robust alternative. The theoretical description of these observables within frameworks like soft-collinear effective theory (SCET) \cite{Bauer:2000yr, Bauer:2001ct, Bauer:2001yt, Bauer:2002nz, Beneke:2002ph} and boosted heavy-quark effective theory (bHQET)~\cite{Fleming:2007qr, Fleming:2007xt} allows for the resummation of large logarithms and a systematic treatment of non-perturbative effects. Such advancements pave the way for extractions of $m_t$ in rigorous short distance schemes.

In addition to precision spectroscopy, the $t\bar t$ system has emerged as a unique laboratory for probing quantum entanglement at the LHC. The top quark's remarkably short lifetime preserves spin correlations against decoherence from hadronization, rendering the $t\bar t$ pair an ideal system for quantum information studies. Both the ATLAS and CMS collaborations recently reported evidence of entanglement in $t\bar t$ events \cite{ATLAS:2023fsd, CMS:2024pts}. Theoretically, observables inspired by quantum information theory are being developed to characterize the entanglement structure of the final state (see, e.g., Refs.~\cite{Han:2023fci, Cheng:2024btk, Dong:2023xiw, Lin:2025eci, Gu:2025ijz}). The boosted regime is particularly advantageous for these measurements, as it enhances the reconstruction of spin correlations and amplifies the sensitivity to the underlying QCD dynamics. A precise theoretical characterization of boosted top-quark production is therefore a crucial prerequisite for fully realizing the diverse physics potential of the LHC.

\begin{figure}[t]
    \centering
    \includegraphics[width=10cm]{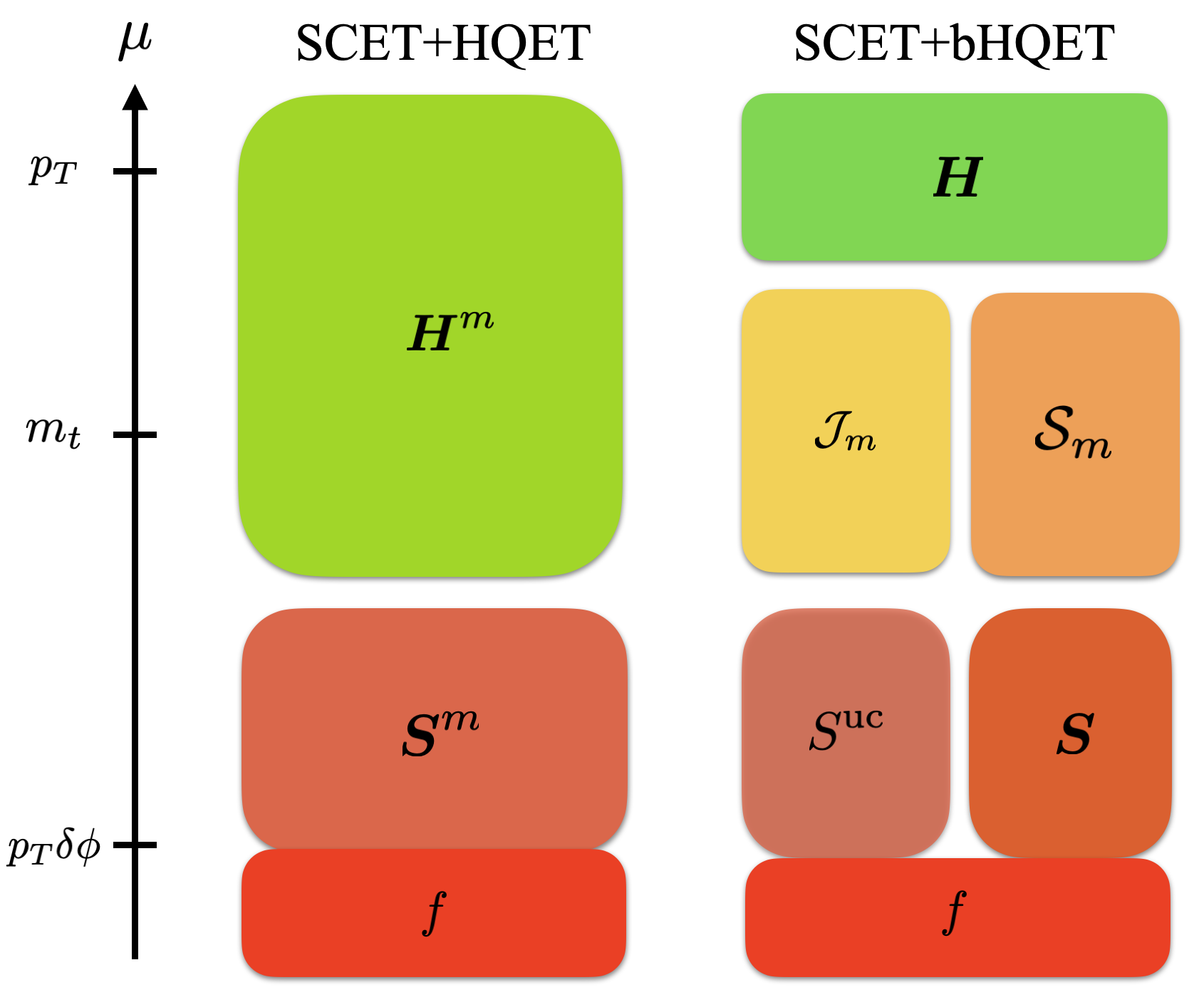}
    \caption{A schematic illustration of the two-step factorization procedure. First, full QCD is matched onto a hybrid SCET and HQET description, where $\boldsymbol{H}^m$  denotes the massive hard function, $\boldsymbol{S}^m$ the massive TMD soft function, and $f$ the standard transverse-momentum-dependent parton distribution functions. Second, in the boosted limit, the HQET sector is matched onto bHQET, introducing the massless hard function $\boldsymbol{H}$, the massive jet function $\mJ_m$, the massive soft function $\mS_m$, the massless TMD soft function $\boldsymbol{S}$, and the ultra-collinear function $S^{\text{uc}}$.} 
    \label{fig:fac}
\end{figure}

Exploiting the full scope of these physics opportunities requires a rigorous theoretical characterization of boosted top quark production. Originally introduced for $e^+e^-$ event shapes with massive quarks~\cite{Fleming:2007qr, Fleming:2007xt}, bHQET has been advanced to $\text{N}^3\text{LL}$ accuracy for lepton colliders~\cite{Gritschacher:2013tza, Pietrulewicz:2014qza, Hoang:2019fze, Bris:2020uyb, Bachu:2020nqn, Bris:2024bcq} and applied to threshold resummation for top pair production at hadron colliders up to NNLL$^\prime$ precision~\cite{Ferroglia:2012ku, Pecjak:2016nee, Czakon:2018nun}. In this work, we present a key development in this direction by deriving a transverse-momentum-dependent (TMD) factorization and resummation framework for back-to-back $t\bar{t}$ production within the kinematic hierarchy $p_T \gg m_t \gg p_T \delta\phi$, where $p_T$ and $\delta\phi$ represent the transverse momentum of the top quark and the azimuthal decorrelation of the $t\bar t$ pair, respectively. To systematically resum the two classes of large logarithms, $\ln\delta\phi$ and $\ln(m_t/p_T)$, we employ a two-step effective field theory matching procedure, as illustrated in Fig.~\ref{fig:fac}. First, degrees of freedom with virtualities of order $p_T \sim m_t$ are integrated out from QCD, resulting in a hybrid effective theory combining SCET for the initial-state radiation with heavy-quark effective theory (HQET) for the final-state top quark system. Second, in the boosted limit $p_T \gg m_t$, the HQET description of the top quarks is matched onto bHQET. A key novel contribution of this work is the derivation of the two-loop ultra-collinear function, which describes soft radiation collinear to the boosted heavy quarks. We extract this function by refactorizing the fully differential massive soft function, thereby completing the set of perturbative inputs required for $\text{NNLL}^\prime$ resummation. The resulting framework establishes a new benchmark for QCD resummation in the boosted regime, offering improved theoretical control for precision top-quark studies and new physics searches at the LHC.

The remainder of this paper is organized as follows. Sec.~\ref{sec:fac} details the factorization formalism in both $\mathrm{SCET}\,+\,\mathrm{HQET}$ and $\mathrm{SCET}\,+\,\mathrm{bHQET}$. In Sec.~\ref{sec:summary of fac}, we review the known two-loop ingredients for the factorization formula and derive the unknown two-loop ultra-collinear function. We present our numerical predictions at NNLL$^\prime$ accuracy in Sec.~\ref{sec:num} and conclude in Sec.~\ref{sec:Conclusions}.

\section{Kinematics and factorization}\label{sec:fac}
We consider the production of top-quark pairs in proton-proton collisions, labeling the four relevant particles in the initial and final states as $a$, $b$, $c$, and $d$
\begin{equation}
    N(P_a) + N(P_b) \rightarrow t(p_c) + \Bar{t}(p_d) + X(p_X),
\end{equation}
where $N(P_a)$ and $N(P_b)$ denote the two colliding protons, \(X\) denotes an inclusive hadronic remnant. At the partonic level, leading-order (LO) contributions arise from two distinct subprocesses: the quark-antiquark annihilation channel,
\begin{equation}
    q(p_a) + \Bar{q}(p_b) \rightarrow t(p_c) + \Bar{t}(p_d),
\end{equation}
and the gluon-gluon fusion channel,
\begin{equation}
    g(p_a) + g(p_b) \rightarrow t(p_c) + \Bar{t}(p_d),
\end{equation}
with incoming parton momenta parameterized as \(p_a^\mu=x_a P_a^\mu\) and \(p_b^\mu=x_b P_b^\mu\). To describe the kinematics of the process, we introduce the Mandelstam variables and Bjorken scaling variables, commonly used in the analysis of $pp$ collisions
\begin{equation}
\begin{aligned}
    &S = (P_a + P_b)^2, \qquad x_a = \frac{p_T}{\sqrt{S}}(e^{y_c} + e^{y_d}), \qquad x_b = \frac{p_T}{\sqrt{S}}(e^{-y_c} + e^{-y_d}),  \\
    &M^2=\Hat{s} = (p_a + p_b)^2 = x_ax_bS, \qquad \Hat{t} = -p_T^2(1 + e^{y_d - y_c}),\qquad \Hat{u} = -p_T^2(1 + e^{y_c - y_d}).
\end{aligned}
\end{equation}
Here, $y_c$ and $y_d$ denote the rapidities of the outgoing quark and antiquark, respectively, while $p_T$ represents the transverse momentum of the final-state particles. 

At LO, the top quark pair is exactly back-to-back in azimuth, \(\Delta\phi\equiv \phi_c-\phi_d=\pi\). We therefore define the small decorrelation angle \(\delta\phi\equiv \pi-\Delta\phi\), which is generated by radiation recoiling out of the scattering plane. Choosing the scattering plane as the \(yz\)-plane (with the beam along \(z\)), the out-of-plane recoil is carried by the \(x\)-component of the transverse momentum imbalance, $q_x$. In the back-to-back limit \(\delta\phi\to 0\), we have \( |q_x| = p_T\,\delta\phi \).

\subsection{Factorization in HQET}
Within HQET and  SCET, factorization and resummation for $t\bar{t}$ production in the threshold limit have been derived in several standard implementations, including pair-invariant-mass and one-particle-inclusive kinematics~\cite{Ahrens:2010zv, Ahrens:2011mw}. Recently the factorization of 0-jettiness $\mathcal{T}_0$ for $t\bar{t}$ production and its NNLL resummation were established in~\cite{Alioli:2021ggd}. Furthermore, SCET provides a TMD factorization and resummation framework for \(t\bar t\) pairs~\cite{Zhu:2012ts, Li:2013mia, Ju:2024xhd}, which was also developed in the language of the perturbative QCD framework~\cite{Catani:2018mei}.

In this section, we review the TMD factorization in an HQET-based formulation, appropriate for the hierarchy $p_T \sim m_t \gg p_T\delta\phi$.  We organize the dynamics in terms of modes with the following scalings in light-cone components
\begin{align}
\begin{split}
    \textbf{hard}: & ~~p_{\rm h}^\mu \sim p_T \,(1,1,1), \\
    n_{a,b}\text{-}\textbf{collinear}:&~~p_{\rm c}^\mu \sim p_T\,(\delta\phi^2,1,\delta\phi), \\
    \textbf{soft}:&~~p_{\rm s}^\mu\sim p_T\,(\delta \phi, \delta \phi ,\delta \phi).
\end{split}
\end{align}
We introduce two light-like reference vectors aligned with the beam directions in Minkowski space,
\begin{equation}
    n^\mu = (1,0,0,1), \qquad \Bar{n}^\mu = (1,0,0,-1),
\end{equation}
so that any four-momentum decomposes as $p^\mu = \left( n\cdot p, \Bar{n} \cdot p , p^\mu_\perp \right) $. 
The vectors $n_i^\mu$ correspond to the light-like directions associated with the initial-state proton beams ($n_{a,b}$). 
For the outgoing heavy quarks we separate the hard kinematics from the residual soft recoil by writing

\begin{equation}
    p_{Q,i}^\mu=m_tv_i^\mu+k_i^\mu,
\end{equation}
where \(v_i^\mu=v_{c,d}^\mu\) are the heavy-quark four velocities with \(v_i^2=1\), and the residual momenta \(k_i^\mu=k_{c,d}^\mu\) scale as \(k_i \sim p_T\,\delta\phi\), represent soft fluctuations. After the standard matching from QCD onto SCET$+$HQET, the effective Hamiltonian for $(q\bar{q}, gg \rightarrow t\bar{t})$ takes the form~\cite{Ahrens:2010zv}
\begin{align}\label{eq:HQET}
    \mathcal{H}_{\rm{HQET}}= \sum_{I,u}\int \d t_1 \d t_2 e^{i m_t (v_c+v_d) \cdot x}\left[C_{Iu}^{q\bar{q}}(t_1,t_2)O_{Iu}^{q\bar{q}}(x,t_1,t_2)+C_{Iu}^{gg}(t_1,t_2) O_{Iu}^{gg}(x,t_1,t_2)\right],
\end{align}
where $I$ labels the color structures, $u$ labels the Dirac structures, and the Wilson coefficients $C_{Iu}^{q\bar q, \, gg}$ encode the hard matching. The relevant effective operators are
\begin{align}
    O_{Iu}^{q\bar{q}}(x,t_1,t_2)&=\sum_{\{a\}}\left(c_I^{q\bar{q}}\right)_{\{a\}}\bar{\chi}_{n_b}^{a_2}(x+t_2\bar{n}_b)\Gamma_u^{\prime}\chi_{n_a}^{a_1}(x+t_1\bar{n}_a)\bar{h}_{v_c}^{a_3}(x)\Gamma_u^{\prime\prime}h_{v_d}^{a_4}(x),\\
    O_{Iu}^{gg}(x,t_1,t_2)&=\sum_{\{a\}}\left(c_I^{gg}\right)_{\{a\}}\mathcal{A}_{n_a\rho\perp}^{a_1}(x+t_1\bar{n}_a)\mathcal{A}_{n_b\sigma\perp}^{a_2}(x+t_2\bar{n}_b)\bar{h}_{v_c}^{a_3}(x)\Gamma_u^{\rho\sigma}h_{v_d}^{a_4}(x),
\end{align}
where \(\Gamma_u^{\prime}\) and \(\Gamma_u^{\prime\prime}\) (and \(\Gamma_u^{\rho\sigma}\)) form a linearly independent basis in Dirac space, and the external vectors \(n_a,n_b,v_c,v_d\) are aligned with the corresponding external legs.  The gauge-invariant SCET quark and gluon fields are denoted by \(\chi_{n_i}\) and \(\mathcal{A}_{n_i\perp}\). The HQET fields $h_{v_i}$ are defined by
\begin{equation}
    h_{v_i}(x) \equiv e^{im_tv_i \cdot x} \,\frac{1 + \slashed{v}_i}{2}\, Q(x),
\end{equation}
where $Q(x)$ represents the heavy-quark field in full QCD. The color tensors \(c_I\) define a basis in color space, and their indices \(\{a\}\equiv\{a_1,a_2,a_3,a_4\}\) label the colors of the external partons. For each channel, we adopt singlet-octet bases as detailed in Sec.~\ref{hard}.

At the leading power, the operators $O_{Iu}$ further factorize into products of collinear fields, heavy-quark fields, and soft Wilson line operators. The resulting operators in the quark annihilation channel take the form as
\begin{equation}
    O^{q\bar q}_{Iu}(x,t_1,t_2)=\sum_{\{a\},\{b\}}\left(c_I\right)_{\{a\}}[O_{m,u}^h(x)]^{b_3b_4}\left[O_u^c(x,t_1,t_2)\right]^{b_1b_2}\left[\boldsymbol{O}^s_m(x)\right]^{\{a\},\{b\}},
\end{equation}
where
\begin{equation}\label{massive operator}
\begin{aligned}
    [O_{m,u}^h(x)]^{b_3b_4}&=\bar{h}_{v_c}^{b_3}(x)\,\Gamma_u^{\prime\prime} \, h_{v_d}^{b_4}(x),\quad\left[O_u^c(x,t_1,t_2)\right]^{b_1b_2}=\bar{\chi}_{n_b}^{b_2}(x+t_2\bar{n}_b)\,\Gamma_u^{\prime}\,\chi_{n_a}^{b_1}(x+t_1\bar{n}_a),\\
    &\left[\boldsymbol{O}^s_m(x)\right]^{\{a\},\{b\}}=[\boldsymbol{S}_{v_c}^\dagger(x)]^{b_3a_3}\left[\boldsymbol{S}_{v_d}(x)\right]^{a_4b_4}[\boldsymbol{S}_{n_b}^\dagger(x)]^{b_2a_2}\left[\boldsymbol{S}_{n_a}(x)\right]^{a_1b_1}.
\end{aligned}
\end{equation}
The soft Wilson lines along the light-like vector $n_a$ and time-like vector $v_c$ are given by
\begin{align}
    &\boldsymbol{S}_{n_a}(x)=\mathbf{P} \exp \left[i g_s \int_{-\infty}^0 \md s \, n_a \cdot \boldsymbol{A}_{s}(x+s n_a)\right],\nonumber\\
    &\boldsymbol{S}_{v_c}(x)=\mathbf{P} \exp \left[-i g_s \int_{0}^\infty \md s \, v_c \cdot \boldsymbol{A}_{s}(x+s v_c)\right],
\end{align}
and analogously for the Wilson lines along $n_b$ and $v_d$. Here $\mathbf{P}$ denotes the path ordering. The integration limits implement the standard convention of past-pointing Wilson lines for incoming partons and future-pointing Wilson lines for outgoing heavy quarks. In the gluon fusion channel, one replaces the collinear quark operators and Dirac structures by their gluonic counterparts and promotes the Wilson lines \(\boldsymbol{S}_{n_{a,b}}\) to the adjoint representation.

Building upon previous studies \cite{Zhu:2012ts, Li:2013mia, Ju:2024xhd}, the TMD factorization formula can be written in momentum space as
\begin{align}\label{hqetcross section}
  \frac{\mathrm{d}^4 \sigma}{\mathrm{d}y_c\,\mathrm{d}y_d\, \mathrm{d} p_T^2\,  \mathrm{d} q_x } = & \sum_{a b} \frac{1}{16 \pi \hat{s}^2} \int_{\perp} x_a f_a\left(x_a, k_{ax}, \mu, \zeta_{a}/\nu^2\right) x_b f_b\left(x_b, k_{bx}, \mu, \zeta_{b}/\nu^2\right) \notag \\ 
 & \times \operatorname{Tr}\left[\boldsymbol{S}_{a b \rightarrow c d}^{m}\left(\lambda_x,M, m_t, y^*,\mu, \nu\right)  \boldsymbol{H}^m_{a b \rightarrow c d}\left( M, m_t, y^*, \mu\right)\right] ,
\end{align}
where $y^*\equiv(y_c-y_d)/2$, and the sum runs over the flavors of the initial state partons $a,b$. The integration measure is defined as
\begin{align}
    \int_{\perp}\equiv\int \mathrm{d} k_{ax} \, \mathrm{d} k_{bx} \, \mathrm{d} \lambda_{x} \,\delta\left(k_{ax}+k_{bx}+\lambda_{x}-q_x\right).
\end{align}
The hard function $\boldsymbol{H}^m_{a b \rightarrow c d}$  is a matrix in color space and encodes the finite part of the virtual corrections at the hard scale $M$. Up to next-to-leading order (NLO) the relevant helicity and color amplitudes can be extracted from \textsc{Recola} \cite{Actis:2012qn, Actis:2016mpe}, and grid-based next-to-next-to-leading order (NNLO) results are available in \cite{Chen:2017jvi}, with further progress toward analytic evaluations reported in \cite{DiVita:2019lpl, Badger:2021owl, Mandal:2022vju, Wang:2023qbf}. The trace $\operatorname{Tr}$ is taken in color space and yields the color-singlet cross section. $f_{a,b}$ represents the unsubtracted transverse-momentum-dependent parton distribution function (TMDPDF) for the incoming partons with flavors $a$ and $b$. For perturbative transverse momentum, they can be matched onto standard PDFs, and the matching coefficients are known to two loops in~\cite{Gehrmann:2012ze, Gehrmann:2014yya, Echevarria:2016scs, Lubbert:2016rku, Luo:2019bmw, Luo:2019hmp} and three loops in~\cite{Luo:2019szz, Luo:2020epw, Ebert:2020qef, Ebert:2020yqt}. In Eq.~\eqref{hqetcross section}, $\mu$ and $\nu$ denote the renormalization and rapidity scales, while $\zeta_{a,b}$ are Collins-Soper parameters~\cite{boussarie2023tmd, Collins:2011zzd}. We will adopt the standard Collins-Soper-Sterman (CSS) formalism \cite{Collins:1981uk, Collins:1984kg} to resum the rapidity logarithms associated with the separation of $n_{a,b}$-collinear and soft modes.

The massive TMD soft function in HQET is defined by the vacuum matrix element of the soft Wilson-line operator in Eq.~\eqref{massive operator},
\begin{equation}
\boldsymbol{S}_{a b \rightarrow c d}^{m}\left(\lambda_x, M, m_t, y^*, \mu, \nu\right)=\int \frac{\mathrm{d} b_x}{2 \pi} e^{i \lambda_x b_x}\langle 0|\overline{\mathbf{T}}[{\boldsymbol{O}^{s\dagger}_m}\left(b^\mu\right)] \mathbf{T}\left[\boldsymbol{O}^s_m(0)\right]| 0\rangle,
\end{equation}
where $b^\mu = (0, b_x, 0, 0)$, and $\lambda_x$ represents the $x$-component of the transverse momentum for soft radiation. The operator $\mathbf{T}(\overline{\mathbf{T}})$ represents (anti-)time ordering. Following the standard color-space formalism~\cite{Catani:1996jh, Catani:1996vz}, we can express the matrix form of the soft function as
\begin{equation}
\boldsymbol{S}_{a b \rightarrow c d, I J}^{m}\left(\lambda_x,M, m_t, y^*, \mu, \nu\right)=\left\langle c_I\left|\boldsymbol{S}_{a b \rightarrow c d}^{m}\left(\lambda_x, M, m_t, y^*, \mu, \nu\right)\right| c_J\right\rangle,
\end{equation}
where $|c_{I,J}\rangle$ denote the basis vectors in color space. This soft function can be derived from the fully differential TMD soft function for top quark pair production at hadron colliders. The azimuthally averaged TMD soft functions were first obtained at NLO in~\cite{Zhu:2012ts, Li:2013mia} and subsequently extended to NNLO in~\cite{Angeles-Martinez:2018mqh, Catani:2023tby}. For the fully differential case, the NLO results have been established in Refs.~\cite{Catani:2014qha, Catani:2021cbl, Ju:2022wia}. More recently, the fully differential soft function at NNLO has been derived for $Q\bar{Q}V$ production at lepton colliders~\cite{Liu:2024hfa}, with the corresponding hadronic results provided in Ref.~\cite{Liu:2025ldi}.

Since the factorization formula in Eq.~\eqref{hqetcross section} involves momentum-space convolutions, we perform a Fourier transform to impact-parameter space ($b$-space) conjugate to $q_x$
\begin{align}
    f_a\left(x_a, b_x, \mu, \zeta_{a}/\nu^2\right)&= \int \mathrm{d}k_{ax} \,e^{-ik_{ax}b_x} f_a\left(x_a, k_{ax}, \mu, \zeta_{a}/\nu^2\right),\\
    \boldsymbol{S}_{a b \rightarrow c d}^{m}\left(b_x, M,m_t,y^*, \mu, \nu\right)&= \int \mathrm{d}\lambda_{x}\, e^{-i\lambda_{x}b_x}\boldsymbol{S}_{a b \rightarrow c d}^{m}\left(\lambda_x,M, m_t, y^*,\mu, \nu\right).
\end{align}
The cross section then takes the compact form
\begin{align}\label{eq:HQET_fac_bspace}
  \frac{\mathrm{d}^4 \sigma}{\mathrm{d}y_c\,\mathrm{d}y_d\, \mathrm{d} p_T^2\,  \mathrm{d} q_x } &=  \sum_{a b} \frac{1}{16 \pi \hat{s}^2} \int \frac{\mathrm{d}b_x}{2\pi}e^{ib_xq_x} x_a f_a\left(x_a, b_x, \mu, \zeta_{a}/\nu^2\right)x_bf_b\left(x_b, b_x, \mu, \zeta_{b}/\nu^2\right)\notag \\
  &\times \operatorname{Tr}\left[\boldsymbol{S}_{a b \rightarrow c d}^{m}\left(b_x, M,m_t, y^*,\mu, \nu\right)  \boldsymbol{H}^m_{a b \rightarrow c d}\left( M, m_t, y^*, \mu\right)\right].
\end{align}
Here we work within the standard TMD factorization assumption and neglect Glauber modes, which could induce factorization-breaking effects in hadronic processes~\cite{Collins:2007nk, Rogers:2010dm, Catani:2011st, Forshaw:2012bi}. The potential size of such effects in the present kinematics remains an open question that can be quantified through dedicated phenomenological studies and, ultimately, comparisons with future high precision measurements.

\subsection{Factorization in bHQET}

Boosted heavy-quark effective theory (bHQET) was introduced in Refs.~\cite{Fleming:2007qr, Fleming:2007xt} in the context of $e^+e^-$ event shapes with massive quarks. Two-loop results are now available for the corresponding bHQET ingredients, enabling resummation up to $\text{N}^3\text{LL}$ accuracy in this setting, see Refs.~\cite{Gritschacher:2013tza, Pietrulewicz:2014qza, Hoang:2019fze, Bris:2020uyb, Bachu:2020nqn, Bris:2024bcq}. The bHQET factorization has also been applied to threshold resummation for top-quark pair production in hadronic collisions~\cite{Ferroglia:2012ku}, and developed up to NNLL$^\prime$ accuracy~\cite{Pecjak:2016nee, Czakon:2018nun}.  The bHQET Lagrangian can be obtained in two equivalent ways. One starts from the $n$-collinear sector of the $\text{SCET}_\text{M}$  Lagrangian~\cite{Leibovich:2003jd, Rothstein:2003wh, Chay:2005ck} and integrates out particle modes with (hard-)collinear virtuality $p^2 \sim m_t^2$~\cite{Dai:2021mxb, Kang:2020xgk, delCastillo:2020omr, delCastillo:2021znl}. The other begins with the HQET Lagrangian and performs a systematic expansion of the heavy-quark field in the boosted limit~\cite{Beneke:2023nmj}. In this work we follow the latter approach to derive the factorization formula used below.

In the presence of a heavy quark, one typically matches the $n_f=n_l+1$ flavor theory onto an effective theory with $n_l$ active flavors as the renormalization scale crosses the heavy-quark threshold. Heavy-quark effects are integrated out and absorbed into matching coefficients so that physical observables are reproduced order by order in perturbation theory. In the $\overline{\mathrm{MS}}$ scheme, the matching of the strong coupling across the top-quark threshold follows from virtual heavy-quark loop corrections and reads
\begin{align}\label{eq:heavy quark decoupling}
     \alpha_s^{(n_f)}(\mu) = \alpha_s^{(n_l)}(\mu)\left[ 1+\frac{4}{3}T_F \frac{ \alpha_s^{(n_l)}(\mu) }{4\pi}\ln{ \frac{\mu^2}{m_t^2} } +\oat \right].
\end{align}

In the kinematic regime $M\sim p_T \gg m_t \gg |q_x|$, the scale separation also implies an additional refactorization of Eq.~\eqref{eq:HQET_fac_bspace}. In particular, the massive hard function in Eq.~\eqref{eq:HQET_fac_bspace} can be factorized as
\begin{equation}
\begin{aligned}\label{Hm}
    \boldsymbol{H}^m_{a b \rightarrow c d}\left( M, m_t, y^*, \mu\right) &= \boldsymbol{H}_{a b \rightarrow c d}\left( M, y^*, \mu\right)\\
    &\hspace{-1cm}\times\mathcal{J}_c\left(m_t,\mu,{\nu}\right)\mathcal{J}_d\left(m_t,\mu,{\nu}\right)\mathcal{S}_m\left(m_t,\mu,\nu\right)\left[1+\mathcal{O}\left(\frac{m_t}{M}\right)\right],
\end{aligned}
\end{equation}
where $\boldsymbol{H}_{a b \rightarrow c d}\left( M, y^*, \mu\right)$ is the massless $2 \to 2$ hard function at the scale $\mu \sim M$. The functions $\mathcal{J}_{c,d}$ and $\mathcal{S}_m$ encode the dynamics of the collinear and soft mass modes at the scale \(\mu\sim m_t\). They can be obtained from the leading term in the small-mass (high-energy) expansion of the massive quark form factor, which is known up to three loops~\cite{Fael:2022miw}, see also Refs.~\cite{Hoang_2015, Mitov:2006xs}. Fixed-order expressions in the refactorization limit $m_t\ll M$ contain large rapidity logarithms. Within the EFT framework~\cite{Hoang_2015}, these logarithms arise because the relevant collinear and soft mass modes have comparable virtuality while being widely separated in rapidity.

For the low-energy dynamics, the appropriate power counting differs from that of standard HQET. While the latter is typically formulated in the heavy-quark rest frame, where the velocity scales as $v^\mu_i \sim (1, 1, 1)$, it lacks the boost parameter necessary to describe the high-energy regime. In the boosted limit $p_T \gg m_t$, we define the power-counting parameter $\lambda \equiv m_t/p_T \ll 1$ and reorganize the bHQET degrees of freedom accordingly. In this boosted frame, expressed in light-cone components, the heavy-quark reference vectors and residual momenta scale as
\begin{align}
     v_{c,d}^\mu  \sim \left(\lambda, \frac{1}{\lambda},1\right), & \quad k^\mu_{c,d} \sim |q_x|\left(\lambda, \frac{1}{\lambda},1\right).
\end{align}
In this boosted frame, the heavy quark field can be defined as \cite{Dai:2021mxb}
\begin{equation}
    h_{n_c}(x) \equiv \sqrt{\frac{2}{\bar{n}_c \cdot v_c}} e^{im_tv_c \cdot x}\xi_{n_c}(x),
\end{equation}
where $\xi_{n_c}(x)=\frac{\slashed{n}_c \slashed{\bar{n}}_c}{4} Q(x)$ is the collinear field defined as in SCET. The phase factor removes the rapidly oscillating dependence associated with the heavy-quark mass, while the prefactor ensures a boost-invariant normalization. With these building blocks, the operators in Eq.~\eqref{massive operator} can be rewritten in a gauge-invariant bHQET form as
\begin{equation}\label{operator}
    \begin{aligned}
       &\boldsymbol{O}_{m,u}^{h,b} =\bar{h}_{n_c}\boldsymbol{W}_{n_c}\Gamma_u^{\prime\prime}\boldsymbol{W}_{n_d}^\dagger h_{n_d},\\
       &\boldsymbol{O}^{s,b}_m=\boldsymbol{O}^s\boldsymbol{Y}_{v_c}^\dagger \boldsymbol{Y}_{v_d},
    \end{aligned}
\end{equation}
with ${\boldsymbol{O}^s}\equiv\boldsymbol{S}_{n_a} \boldsymbol{S}_{n_b}^{\dagger} \boldsymbol{S}_{n_c}^{\dagger} \boldsymbol{S}_{n_d}$. For brevity we suppress explicit coordinate dependence and color indices.
The Wilson lines $\boldsymbol{W}_{n_i}$ and $\boldsymbol{Y}_{v_i}$ $(i=c,d)$ both describe the additional ultra-collinear modes in bHQET, but have different origins. $\boldsymbol{W}_{n_i}$ comes from the off-shell propagator on the opposite direction external leg, and $\boldsymbol{Y}_{n_i}$ comes from the soft radiations collinear to the quark, which are defined as
\begin{equation}\label{eq:Wilson_lines}
\begin{aligned}
    &\boldsymbol{W}_{n_i}(x)=\mathbf{P} \exp \left[-i g_s \int^{\infty}_0 \md s \, {\bar{n}_i} \cdot \boldsymbol{A}_{\mathrm{uc}}(x+s \bar{n}_i)\right],\\
    &\boldsymbol{Y}_{v_i}(x)=\mathbf{P} \exp \left[-i g_s \int^{\infty}_0 \md s \, {v_i} \cdot \boldsymbol{A}_{\mathrm{uc}}(x+s {v_i})\right].
\end{aligned}
\end{equation}
As a consequence, the massive soft function further factorizes into a massless soft function and ultra-collinear functions, as discussed in Sec.~\ref{uc function}.

Based on the above discussion, we can formulate the bHQET factorization formula in the back-to-back limit. The corresponding momentum mode scales are 
\begin{align}\label{bhqetmode}
\begin{split}
    \textbf{hard}: & ~~p_{\rm h}^\mu \sim p_T \,(1,1,1), \\
    n_{a,b}\text{-}\textbf{collinear}:&~~p_{\rm c}^\mu \sim p_T\,(\delta\phi^2,1,\delta\phi),\\ 
    n_{c,d}\text{-}\textbf{collinear}:&~~p_{\rm c}^\mu \sim p_T\,(\lambda^2,1,\lambda),\\ 
    \textbf{massive soft}:& ~~ p_{\mathrm{ms}}^\mu\sim p_T(\lambda,\lambda,\lambda),\\
    \textbf{soft}:&~~p_{\rm s}^\mu\sim p_T\,(\delta \phi, \delta \phi ,\delta \phi),\\
    n_{c,d}\text{-}\textbf{ultra-collinear}:&~~p_{\mathrm{uc}}^\mu\sim \frac{p_T \delta\phi}{\lambda}\,(\lambda^2,1,\lambda),
\end{split}
\end{align}
where the $n_{a,b}$- and $n_{c,d}$-collinear sectors encode the dynamics of the initial-state beam functions and final-state jet functions, respectively. The $n_{a,b}$-collinear, soft, and $n_{c,d}$-ultra-collinear modes share the same virtuality, $p^2\sim p_T^2\delta\phi^2 \sim q_x^2$, while the $n_{c,d}$-collinear and massive soft modes have virtuality $p^2\sim p_T^2\lambda^2\sim m_t^2$. Since modes with the same virtuality are distinguished by their rapidity, rapidity divergences inherently arise in the factorized description. For the rapidity logarithms associated with the final-state sector, we employ the rapidity renormalization group (RRG) framework of Refs.~\cite{Chiu:2011qc, Chiu:2012ir}, which can equivalently be reproduced using the collinear-anomaly method~\cite{Becher:2010tm}. 
The resulting factorized cross section in momentum space reads
\begin{align}\label{bhqetcross section}
  \frac{\mathrm{d}^4 \sigma}{\mathrm{d}y_c\,\mathrm{d}y_d\, \mathrm{d} p_T^2\,  \mathrm{d} q_x } &=  \sum_{a b} \frac{1}{16 \pi \hat{s}^2} \int_{\perp} x_a f_a\left(x_a, k_{ax}, \mu, \zeta_{a}/\nu^2\right) x_b f_b\left(x_b, k_{bx}, \mu, \zeta_{b}/\nu^2\right) \notag \\ 
 & \times \operatorname{Tr}\left[\boldsymbol{S}_{a b \rightarrow c d}\left(\lambda_{x},y^*, \mu, \nu\right)  \boldsymbol{H}_{a b \rightarrow c d}\left( M, y^*, \mu\right)\right]\mathcal{S}_m\left(m_t,\mu,\nu\right) \notag\\
  & \times \mathcal{J}_c\left(m_t,\mu,{\nu}\right)\mathcal{J}_d\left(m_t,\mu,{\nu}\right)S_c^{\mathrm{uc}}\left(k_{cx}, m_t, \mu ,\nu \right) S_d^{\mathrm{uc}}\left(k_{dx}, m_t, \mu ,\nu \right),
\end{align}
where the sum runs over the flavors of the incoming partons $(a,b)$, $\boldsymbol{S}_{a b \rightarrow c d}$ is the unsubtracted massless TMD soft function which will be discussed in Sec.~\ref{PDF and soft}, and $S^{\mathrm{uc}}_{c,d}$ are the TMD ultra-collinear functions describing soft-collinear radiation associated with the boosted heavy quarks at the scale $p_T\delta\phi$. Here $k_{cx}$ and $k_{dx}$ denote the $x$-components of the recoil momenta carried by the ultra-collinear sectors in the $n_c$ and $n_d$ directions. The ultra-collinear functions are analogous to the partonic shape function appearing in threshold factorization for $B$-meson decays~\cite{Gardi:2005yi, Neubert:2007je}. The convolution measure is defined as
\begin{align}
    \int_{\perp} \equiv \int \mathrm{d} k_{ax} \, \mathrm{d} k_{bx} \,\mathrm{d} k_{cx} \,\mathrm{d} k_{dx}\, \mathrm{d} \lambda_{x} \, \delta\left(k_{ax}+k_{bx}+k_{cx}+k_{dx}+\lambda_{x}-q_x\right).
 \end{align}
Performing the Fourier transformation, the factorized cross section then takes the compact form as
\begin{align}\label{cross section}
  \notag \frac{\mathrm{d}^4 \sigma}{\mathrm{d}y_c\,\mathrm{d}y_d\, \mathrm{d} p_T^2\,  \mathrm{d} q_x } &=  \sum_{a b} \frac{1}{16 \pi \hat{s}^2} \int \frac{\mathrm{d}b_x}{2\pi}e^{ib_xq_x} x_a f_a\left(x_a, b_x, \mu, \zeta_{a}/\nu^2\right)
  x_bf_b\left(x_b, b_x, \mu, \zeta_{b}/\nu^2\right)\\
  \notag &\times \operatorname{Tr}\left[\boldsymbol{S}_{a b \rightarrow c d}\left(b_x,y^*, \mu, \nu\right) \boldsymbol{H}_{a b \rightarrow c d}\left( M, y^*, \mu\right)\right]\mathcal{S}_m\left(m_t,\mu,\nu\right) \\
  &\times \mathcal{J}_c\left(m_t,\mu,{\nu}\right)\mathcal{J}_d\left(m_t,\mu,{\nu}\right)S_c^{\mathrm{uc}}\left(b_x,m_t, \mu ,\nu \right) S_d^{\mathrm{uc}}\left(b_x, m_t, \mu ,\nu \right).
\end{align}
Here, all perturbative ingredients required for NNLL$^\prime$ resummation are available, except for the ultra-collinear functions $S^{\mathrm{uc}}_{c,d}$. In the next section, we review the known results and derive the two-loop ultra-collinear functions.

\section{Summary of factorized functions}\label{sec:summary of fac}
In this section, we summarize the perturbative ingredients entering the bHQET factorization formula in Eq.~\eqref{cross section}, along with their renormalization group (RG) evolution equations.

\subsection{Massless hard function}\label{hard}
In the bHQET factorization theorem, the massless hard function \(\boldsymbol{H}_{ab\to cd}\) for process \(ab\to cd\) is the color-space Wilson-coefficient matrix obtained by matching renormalized QCD onto SCET and bHQET at the hard scale \(\mu\sim M\). Equivalently, \(\boldsymbol{H}_{ab\to cd}\) can be constructed from the renormalized \(2\to2\) partonic amplitudes in a chosen color basis (see Ref. \cite{Moult:2015aoa} for a precise definition). The relevant amplitudes are known at one-loop~\cite{Kunszt:1993sd} and two-loop \cite{Anastasiou:2000kg, Anastasiou:2000ue, Glover:2001af, Bern:2002tk, Bern:2003ck, Glover:2003cm, Glover:2004si, DeFreitas:2004kmi}, with recent progress at three-loop~\cite{Bargiela:2021wuy, Caola:2020dfu, Caola:2021izf, Caola:2021rqz}. For the NNLL$^\prime$ accuracy targeted in this work we require \(\boldsymbol{H}_{ab\to cd}\) through NNLO. The NLO hard functions were extracted in Refs.~\cite{Moult:2015aoa, Kelley:2010fn}, and the NNLO hard functions are provided as a \textsc{Mathematica} file in Ref.~\cite{Broggio:2014hoa}, which we use in our numerical implementation.

For the \(q\bar q\to t\bar t\) channel, we employ the singlet-octet color basis
\begin{align}\label{quark color}
    \left(c_1^{q\bar{q}}\right)_{\{a\}} \equiv\left\langle\{a\} \mid c_1^{q\bar{q}}\right\rangle=\delta_{a_1 a_2} \delta_{a_3 a_4}, \quad \left(c_2^{q\bar{q}}\right)_{\{a\}} \equiv\left\langle\{a\} \mid c_2^{q\bar{q}}\right\rangle=t_{a_2 a_1}^c t_{a_3 a_4}^c,
\end{align}
and for the \(gg\to t\bar t\) channel we use
\begin{align}\label{gluon color}
    &\left(c_1^{gg}\right)_{\{a\}} \equiv \left\langle\{a\} \mid c_1^{gg}\right\rangle=\delta^{a_1 a_2} \delta_{a_3 a_4}, \quad   
    \left(c_2^{gg}\right)_{\{a\}} \equiv\left\langle\{a\} \mid c_2^{gg}\right\rangle=i f^{a_1 a_2 c} t_{a_3 a_4}^c, \notag\\
    &\left(c_3^{gg}\right)_{\{a\}} \equiv\left\langle\{a\} \mid c_3^{gg}\right\rangle=d^{a_1 a_2 c} t_{a_3 a_4}^c, 
\end{align}
where $t^a$ are the fundamental generators, and $f^{abc}$ ($d^{abc}$) are the antisymmetric (symmetric) SU(3) structure constants. The indices $a_{1,2}$ refer to the incoming partons, while $a_{3,4}$ refer to the outgoing heavy quarks.

The hard function obeys the RG equation
\begin{equation}
\frac{\mathrm{d}}{\mathrm{d} \ln \mu} \boldsymbol{H}=\boldsymbol{\Gamma}_H \boldsymbol{H}+\boldsymbol{H} \, \boldsymbol{\Gamma}_H^{\dagger},
\end{equation}
with the anomalous dimension
\begin{equation}\label{hardad}
    \boldsymbol{\Gamma}_{H}=\left[\frac{C_H}{2} \gamma_{\text {cusp }}^{\nf}\left(\mu\right)\left(\ln \frac{M^2}{\mu^2}-i \pi\right)+\gamma_H^{\nf}\left(\mu\right)\right] \mathbf{1}+\gamma_{\text {cusp }}^{\nf}\left(\mu\right) \boldsymbol{M}_{a b \rightarrow c d},
\end{equation}
where $C_H=n_q C_F + n_g C_A$ and $\gamma_H = n_q\gamma_q + n_g\gamma_g$. Here $\gamma_{q,g}$ denote the quark and gluon anomalous dimensions, and \(n_q\) (\(n_g\)) is the number of external quark (gluon) legs. The non-diagonal contribution is defined as
\begin{equation}
\boldsymbol{M}_{a b \rightarrow c d}=(\ln r+i \pi) \boldsymbol{M}_{1, a b \rightarrow c d}+\ln \frac{r}{1-r} \boldsymbol{M}_{2, a b \rightarrow c d},
\end{equation}
with \(r\equiv-\hat{t}/\hat{s}\). Explicit expressions for the matrices $\boldsymbol{M}_{1,2}$ in these color bases are given in Ref.~\cite{Broggio:2014hoa}.

Our NNLL$^\prime$ resummation incorporates the cusp anomalous dimension at three loops and the non-cusp anomalous dimensions at two loops. The required anomalous dimensions are collected in Appendix~\ref{app:a-dim}. At this order, the dipole structure in Eq.~\eqref{hardad} is sufficient. Non-dipole color kinematic correlations, which first appear at three loops, enter beyond the precision considered here~\cite{Almelid:2015jia, Almelid:2017qju}. For a comprehensive account of earlier studies on the structure and color space evolution in Eq.~\eqref{hardad}, see Refs.~\cite{Kidonakis:1998nf, Dixon:2008gr, Gardi:2009qi, Gardi:2009zv, Dixon:2009ur, Becher:2009cu, Becher:2009qa}.

\subsection{Massive jet and soft functions}

In this subsection, we detail the refactorized massive jet functions $\mathcal{J}_{c,d}$ and the massive soft function $\mathcal{S}_m$ introduced in Eq.~\eqref{Hm}. These functions appear as matching coefficients at the threshold scale $\mu \sim m_t$, effectively bridging the $n_f$-flavor (high-energy) and $n_l$-flavor (low-energy) theories. They encapsulate the finite top-quark mass corrections associated with the collinear and soft sectors.

At one-loop order, the massive jet function is given by
\begin{equation}
\mJ_{i}^{(1)}(m_t, \mu, \nu)=\frac{\alpha_s^{\nl}}{4 \pi} C_F\left[\left(1+\ln \frac{\mu^2}{m_t^2}\right) \ln \frac{\mu^2}{m_t^2}+4+\frac{\pi^2}{6}\right],
\end{equation}
while the one-loop massive soft function vanishes, $\mathcal{S}_m^{(1)} = 0$. Starting at two loops, the hierarchy $M \gg m_t$ induces additional large logarithms within the massive hard function that cannot be resummed via standard RG evolution in $\mu$. In the EFT framework, these logarithms originate from the fact that the massive soft and \(n_{c,d}\)-collinear sectors share a comparable virtuality ($\sim m_t$) but are widely separated in rapidity. To systematically resum these logarithms, the massive hard function must be refactorized as shown in Eq.~\eqref{Hm}. Explicit expressions for $\mathcal{J}_i$ and $\mathcal{S}_m$ up to two loops are provided in Ref.~\cite{Hoang_2015}.
 
The functions \(\mathcal{J}_i\) and \(\mathcal{S}_m\) satisfy RG equations in the scale \(\mu\),
\begin{align}
\notag \frac{\mathrm{d}}{\mathrm{d} \ln \mu} \mJ_i\left(m_t,\mu,\nu\right)& = \Gamma^{\mJ_{i}}\left(\mu\right)\mJ_i\left(m_t,\mu,\nu\right),\\
\frac{\mathrm{d}}{\mathrm{d} \ln \mu} \mS_m\left(m_t,\mu,\nu\right)& = \Gamma^{\mS_{m}}\left(\mu\right)\mS_m\left(m_t,\mu,\nu\right),
\end{align}
with the anomalous dimensions
\begin{align}
\notag \Gamma^{\mJ_i}\left(\mu\right)&=-C_F \gamma_{\text {cusp }}^{\nl}\left(\mu\right) \ln \frac{m_t^2}{\mu^2}+\gamma^{\nl}_{\mJ_m}(\mu)+\gamma^{\text{Ex},\nl}_{\mJ_i}(\mu),\\
\Gamma^{\mS_m}\left(\mu\right)&=\gamma^{ \text{Ex},\nl}_{\mS_m}(\mu).
\end{align}
The ``Ex'' terms arise from the heavy quark decoupling relation in Eq.~(\ref{eq:heavy quark decoupling}), and are given by 
\begin{align}
    &\notag \gamma^{\text{Ex},\nl}_{\mJ_i}(\mu)=\left(\frac{\as^\nl}{4\pi}\right)^2 \cf T_F \left[-16\left(\frac{1}{3}\ln\frac{m_t^2}{\mu^2}+\frac{5}{9}\right)\ln\frac{\nu^2}{\omega_i^2}-8\ln\frac{m_t^2}{\mu^2}-\frac{4}{3}-\frac{16\pi^2}{9}\right],\\
    &\gamma^{\text{Ex},\nl}_{\mS_m}(\mu)=\left(\frac{\as^\nl}{4\pi}\right)^2 \cf T_F\left[32\left(\frac{1}{3}\ln\frac{m_t^2}{\mu^2}+\frac{5}{9}\right)\ln\frac{\nu^2}{\mu^2}-\frac{448}{27}+\frac{8\pi^2}{9}\right],
\end{align}
where \(\omega_i=\bar{n}_i\cdot p_i\) denotes the large light-cone momentum component of the heavy quark.

In addition to the standard RG evolution, \(\mathcal{J}_i\) and \(\mathcal{S}_m\) obey RRG equations in the scale \(\nu\),
\begin{align}
    \notag \frac{\mathrm{d}}{\mathrm{d} \ln \nu} \mJ_i\left(m_t,\mu,\nu\right)& = \Gamma^{\mJ_{m}}_\nu\left(\mu\right)\mJ_i\left(m_t,\mu,\nu\right),\\
    \frac{\mathrm{d}}{\mathrm{d} \ln \nu} \mS_m\left(m_t,\mu,\nu\right)& = \Gamma^{\mS_{m}}_\nu\left(\mu\right)\mS_m\left(m_t,\mu,\nu\right),
\end{align}
with rapidity anomalous dimensions
\begin{align}
    \notag \Gamma^{\mJ_{m}}_\nu(\mu) 
    &=-2 \cf \left[A^{(n_f)}_\Gamma(m_t,\mu)-A^{(n_l)}_\Gamma(m_t,\mu)\right]+\Gamma_\nu^{\mJ_m}(m_t), \\ 
    \Gamma^{\mS_{m}}_\nu(\mu) &=-2\Gamma^{\mJ_{m}}_\nu(\mu),
\end{align}
and the two-loop boundary condition
\begin{align}
    \Gamma_\nu^{\mJ_m}(m_t)=\left(\frac{\as^{(n_l)}}{4\pi}\right)^2\cf T_F \frac{224}{27}.
\end{align}
Here \(A_\Gamma^{(n_i)}\) is the standard evolution function with \(n_i\) active flavors,
\begin{align}\label{eq:Agamma}
    A^{(n_i)}_\Gamma(\mu_1,\mu_2)=-\int_{\mu_1}^{\mu_2}\frac{\d \bar{\mu}}{\bar{\mu}}\gamma_{\text{cusp}}^{(n_i)}(\bar{\mu}).
\end{align}

\subsection{TMDPDFs and soft function}\label{PDF and soft}
In this subsection, we outline the Collins-Soper treatment of rapidity divergences and specify the TMDPDF conventions used in our analysis. To define TMDPDFs free of rapidity divergences, we adopt the Collins-Soper scheme~\cite{boussarie2023tmd, Collins:2011zzd}, in which a universal Drell-Yan soft function \(S_{ab}(b_x,\mu,\nu)\) is absorbed into the unsubtracted beam functions. Specifically, the subtracted TMDPDFs are defined as
\begin{equation}\label{css}
\begin{aligned}
f_{a / p}\left(x_a, b_x, \mu, \zeta_a / \nu^2\right)& f_{b / p}\left(x_b, b_x, \mu, \zeta_b / \nu^2\right)S_{a b}(b_x, \mu, \nu) \\
 &\equiv \tilde f_{a / p}\left(x_a, b_x, \mu, \zeta_a\right) \tilde f_{b / p}\left(x_b, b_x, \mu, \zeta_b\right),
 \end{aligned}
\end{equation}
where $a$ and $b$ denote the color representations of the incoming partons. The subtracted TMDPDFs \(\tilde f_{a/p}\) defined in this way do not carry an explicit \(\nu\)-dependence. Each subtracted TMDPDF obeys the Collins-Soper evolution equation
\begin{equation}
\sqrt{\zeta_a} \frac{\mathrm{d}}{\mathrm{d} \sqrt{\zeta_a}} \tilde f_{a / p}\left(x_a, b_x, \mu, \zeta_a\right)=\kappa_a(b_x, \mu) \tilde f_{a / p}\left(x_a, b_x, \mu, \zeta_a\right),
\end{equation}
where the perturbative Collins-Soper kernel $\kappa_i(b_x, \mu)$ is known up to four loops~\cite{Duhr:2022yyp, Moult:2022xzt}.  The solution of the evolution equation is
\begin{equation}
\tilde f_{a / p}\left(x_a, b_x, \mu, \zeta_{a, f}\right)=\tilde f_{a / p}\left(x_a, b_x, \mu, \zeta_{a, i}\right)\left(\sqrt{\frac{\zeta_{a, f}}{\zeta_{a, i}}}\right)^{\kappa_a(b_x, \mu)}.
\end{equation}
We choose the canonical scales $\zeta_{a,f} = \zeta_{b,f} = \hat{s}$ and $\zeta_{a, i} = \zeta_{b, i} = b_0^2/b_x^2$ to minimize large logarithms at the endpoints of the evolution. Additionally, the RG evolution of the TMDPDFs in $\mu$ is governed by
\begin{equation}
\frac{\mathrm{d}}{\mathrm{d} \ln \mu} \tilde f_{a / p}\left(x_a, b_x, \mu, \zeta_{a, f}\right)=\left[C_a \gamma_{\text {cusp }}^{(n_l)}\left(\mu\right) \ln \frac{\mu^2}{\zeta_{a, f}}-2 \gamma_a^{(n
_l)}\left(\mu\right)\right] \tilde f_{a / p}\left(x_a, b_x, \mu, \zeta_{a, f}\right).
\end{equation}
Here $C_a= \cf$ or $\ca$ is the Casimir of the parton $a$. In the perturbative regime, the TMDPDFs can be matched onto the collinear PDFs via the operator product expansion relation 
\begin{equation}\label{TMDPDF matching}
\begin{aligned}
\tilde f_{a / p}\left(x_a, b_x, \mu, \zeta_{a}\right)  =\sum_i\int\frac{\md y}{y} \mathcal{I}_{ai}\left(\frac{x_a}{y}, b_x, \mu, \zeta_{a}\right) f_{i / p}\left(y, \mu\right)+\mathcal{O}\left(b_x^2\,\Lambda_{\text{QCD}}^2\right),
\end{aligned}
\end{equation}
where \(f_{i/p}(y,\mu)\) is the collinear PDF and the matching coefficients \(\mathcal{I}_{ai}\) are known to two loops~\cite{Gehrmann:2012ze, Gehrmann:2014yya, Echevarria:2016scs, Lubbert:2016rku, Luo:2019bmw, Luo:2019hmp} and three loops~\cite{Luo:2019szz, Luo:2020epw, Ebert:2020qef, Ebert:2020yqt}. For gluons, Lorentz invariance allows the gluon TMDPDFs to be decomposed as
\begin{equation}
    \tilde f_{g / p}^{\mu\nu}\left(x_g,b^\mu,\mu, \zeta_{g}\right) = \frac{g_\perp^{\mu\nu}}{d-2} \tilde f_{g / p}\left(x_g,b^\mu,\mu, \zeta_{g}\right) + \left(\frac{b^\mu b^\nu}{b^2} - \frac{g_\perp^{\mu\nu}}{d-2}\right) \tilde f_{g / p}'\left(x_g,b^\mu,\mu, \zeta_{g}\right),
\end{equation}
where the second term corresponds to the linearly polarized contribution. For the process considered here, the effects of linearly polarized TMDPDFs first enter at $\mathcal{O}(\alpha_s^2)$ as a non-logarithmic term, and we leave a dedicated phenomenological study of these effects to future work.

For the study of azimuthal decorrelation, the relevant global soft function in the bHQET factorization theorem takes a simple form, sharing a similar structure to the TMD soft function studied in Refs.~\cite{Gao:2019ojf, Chien:2020hzh, Chien:2022wiq, Gao:2023ivm, Fu:2024fgj}. Its operator definition is
\begin{equation}\label{soft}
 \boldsymbol{S}_{a b \rightarrow c d}\left(b_x, y^*,\mu, \nu\right)=\langle 0|\overline{\mathbf{T}}[{\boldsymbol{O}^s}^{\dagger}\left(b^\mu\right)] \, \mathbf{T}\left[\boldsymbol{O}^s(0)\right]| 0\rangle,
 \end{equation}
with ${\boldsymbol{O}^s}=\boldsymbol{S}_{n_a} \boldsymbol{S}_{n_b}^{\dagger} \boldsymbol{S}_{n_c}^{\dagger}\boldsymbol{S}_{n_d}$. Here \(n_{a,b}\) align with the incoming beams, and \(n_{c,d}\) are the lightlike reference vectors for the boosted top quarks. The dependence on \(y^*\) arises through the dipole invariants \(n_i\!\cdot n_j\). Using the $\eta$-rapidity regulator, the bare one-loop soft function is given by
\begin{equation}  
    \boldsymbol{S}_{a b \rightarrow c d}^{ (1)}\left(b_x, y^*,\mu,  \nu\right) = -\sum_{i< j}(\mathbf{T}_i \cdot \mathbf{T}_j)\,\,I_{ij}\,,
\end{equation}
with the integral
\begin{align}
    I_{i j}^{\mathrm{bare}} = \frac{\alpha_s^{\nl} \mu^{2 \epsilon} \pi^\epsilon e^{\epsilon \gamma_E}}{\pi^2} \int \d^d k \,\delta\left(k^2\right)&\theta(k^0) e^{i k_x b_x} \frac{n_i \cdot n_j}{\left(n_i \cdot k\right)\left(n_j \cdot k\right)} \left(\frac{\nu}{2k^0}\right)^\eta\,.
\end{align}
Here \(\nu\) is the rapidity scale and \(\mathbf{T}_i\) denotes the color generator acting on the leg \(i\). After renormalization, the one-loop soft function is
\begin{align}
    \boldsymbol{S}_{a b \rightarrow c d}^{(1)}\left(b_x, y^*,\mu, \nu\right)&=\frac{\alpha_s^{\nl}}{4 \pi}\left(-\sum_{i<j}\left(\mathbf{T}_i \cdot \mathbf{T}_j\right)S_\perp^{(1)}\left(L_b, L_\nu - \ln\frac{n_i \cdot n_j}{2}\right)\right)\\
&=\frac{\alpha_s^{\nl}}{4 \pi}\left\{-\sum_{i<j}\left(\mathbf{T}_i \cdot \mathbf{T}_j\right)\left[-2 L_b^2+4 L_b\left(L_\nu-\ln \frac{n_i \cdot n_j}{2}\right)-\frac{\pi^2}{3}\right]\right\},\notag
\end{align}
where $S_\perp^{(n)}$ is the $n$-loop TMD soft function for color-singlet production at hadron colliders, known up to three loops~\cite{Li:2016ctv}. We use the shorthand $L_b = \text{ln}(\mu^2b_x^2/b_0^2)$ with $b_0 = 2e^{-\gamma_E}$, and $L_\nu=\ln (\mu^2/\nu^2)$. At two loops, the soft function contains both dipole and tripole color structures~\cite{Gao:2023ivm},
\begin{equation}\label{2loopsoft}
\begin{aligned}
\boldsymbol{S}_{a b \rightarrow c d}^{(2)}\left(b_x, y^*,\mu, \nu\right) = &-\sum_{i<j} (\mathbf{T}_i \cdot \mathbf{T}_j) S_\perp^{(2)}\left(L_b, L_\nu - \ln\frac{n_i \cdot n_j}{2}\right)
\\
& + \frac{1}{2!} \left[\boldsymbol{S}_{a b \rightarrow c d}^{(1)}\left(b_x, y^*,\mu, \nu\right)\right]^2+ i f^{abc} \mathbf{T}_1^a \mathbf{T}_2^b \mathbf{T}_3^c \, S_{\text{tri}}(b_x, y^*, \mu)
 \,.
\end{aligned}
\end{equation}
Here, the tripole term $S_{\text{tri}}$ is purely imaginary. At the NNLL$^\prime$ accuracy, the two-loop soft function $\boldsymbol{S}_{a b \rightarrow c d}^{(2)}$ contracts with the real tree-level hard function $\boldsymbol{H}_{a b \rightarrow c d}^{(0)}$. Consequently, the purely imaginary tripole term $S_{\text{tri}}$ does not contribute to the physical cross section.

The soft function satisfies the RG equation
\begin{equation}
\frac{\mathrm{d}}{\mathrm{d} \ln \mu} \boldsymbol{S}=\boldsymbol{\Gamma}_S^{\dagger} \boldsymbol{S}+\boldsymbol{S} \, \boldsymbol{\Gamma}_S\,,
\end{equation}
with the anomalous dimension~\cite{Kidonakis:1998bk, Kidonakis:1998nf, Aybat:2006mz, Aybat:2006wq}
\begin{equation}
\boldsymbol{\Gamma}_S = 
\sum_{i<j} \mathbf{T}_i \cdot \mathbf{T}_j \, \gamma_{\text {cusp }}^{(n_l)}(\mu) 
\ln \frac{\sigma_{ij} \nu^2 n_i \cdot n_j - i0}{2\mu^2}
-  \frac{C_H}{2} \gamma_s \mathbf{1}
 \,,
\label{eq:4.3}
\end{equation}
where $\sigma_{ij} = -1$ if both $i$ and $j$ are incoming or both are outgoing, and $\sigma_{ij} = 1$ otherwise.  $\gamma_s$ denotes the threshold soft anomalous dimension~\cite{Li:2014afw}. Consistent with the hard anomalous dimension in Eq.~\eqref{hardad}, we neglect quadrupole color-kinematic correlations, which first arise at three loops. The rapidity evolution equation is
\begin{equation}
    \frac{\md}{\md\ln\nu}\boldsymbol{S} =   \Gamma_S^\nu \, \boldsymbol{S} \, ,
\end{equation}
with 
\begin{equation}
\Gamma_S^\nu = 2C_H\left[2A^{(n_l)}_\Gamma(b_0/b_x,\mu) + \gamma_r^{(n_l)}(b_0/b_x) \right] .
\end{equation}
Here, $\gamma_r$ is the rapidity anomalous dimension for the transverse-momentum distribution~\cite{Li:2016ctv}. Notably, there is no color non-diagonal rapidity anomalous dimension.

\subsection{TMD ultra-collinear function}\label{uc function}

In this subsection, we derive the ultra-collinear functions $S_{c,d}^{\mathrm{uc}}$ and detail our extraction of the two-loop result via refactorization. The ultra-collinear function captures the dynamics of soft-collinear radiation in the final state within the hierarchy $p_T \gg m_t \gg |q_x|$. Its operator definition follows from Eq.~\eqref{operator}, and can be written as
\begin{equation}\label{eq:ucollinear}
S^{\mathrm{uc}}_c(b_x, m_t,  \mu,\nu)=\frac{1}{N_c}\text{Tr}\left\langle0\right|\overline{\mathbf{T}}[\boldsymbol{W}_{n_c}\left(b^\mu\right)\boldsymbol{Y}_{v_c}^\dagger\left(b^\mu\right)]\mathbf{T}[\boldsymbol{Y}_{v_c}\left(0\right)\boldsymbol{W}_{n_c}^\dagger\left(0\right)]|0\rangle ,
\end{equation}
and similarly for $S^{\mathrm{uc}}_d$. Here, \(b^\mu=(0,b_x,0,0)\). The Wilson lines \(\boldsymbol{W}_{n_c}\) (light-like) and \(\boldsymbol{Y}_{v_c}\) (time-like) describe soft-collinear radiation associated with the boosted heavy-quark direction. We parameterize the heavy-quark velocity vector \(v_c^\mu\) as
\begin{equation}
v_c^\mu=\frac{\omega_{c}}{m_t} \frac{n_c^\mu}{2}+\frac{m_t}{\omega_{c}} \frac{\bar{n}_c^\mu}{2},
\end{equation}
with $\omega_{c}=  \bar{n}_c\cdot p_c=2p_T\cosh y_c$. Using the \(\eta\)-rapidity regulator with the measure \((\nu/(k\cdot\bar n_c))^\eta\), the bare one-loop result is
\begin{equation}
\begin{aligned}
S_c^{\mathrm{uc},(1),\mathrm{bare}}&(b_x, m_t, \mu, \nu ) = \frac{\alpha_s^{(n_l)} \mu^{2 \epsilon} \mathrm{e}^{\epsilon \gamma_E}}{2 \pi^{2-\epsilon}} C_F \int \mathrm{d}^d k \delta\left(k^2\right) \theta\left(k^0\right)\left(\frac{\nu}{k \cdot \bar n_c}\right)^\eta \mathrm{e}^{i k_{x} b_{x}} \\
&\hspace{-1cm}\times \left[\frac{2 v_c \cdot \bar n_c}{(v_c \cdot k)(k \cdot \bar n_c)}-\frac{v_c \cdot v_c}{(v_c \cdot k)(k \cdot v_c)}\right]\\
&\hspace{-1cm}= \frac{\alpha_s^{(n_l)} C_F}{4 \pi}\left[\left(\frac{2}{\eta}+\ln \frac{\nu^2}{\zeta_c}\right)\left(\frac{2}{\epsilon}+2 L_b\right)-\frac{2}{\epsilon^2}+\frac{2}{\epsilon}+L_b^2+2 L_b+\frac{\pi^2}{6}\right],
\end{aligned}
\end{equation}
where $\zeta_{c}\equiv\omega_{c}^2 \mu^2/m_t^2$. Renormalizing this result yields
\begin{equation}
    S_c^{\mathrm{uc},(1)}(b_x, m_t, \mu, \nu )=\frac{\alpha_s^{(n_l)} C_F}{4 \pi}\left[2 L_b\ln \frac{\nu^2}{\zeta_{c}}+L_b^2+2 L_b+\frac{\pi^2}{6}\right],
\end{equation}
which agrees with Ref.~\cite{vonKuk:2024uxe}. The rapidity divergence in the ultra-collinear function arises specifically from the presence of the light-like Wilson line \(\boldsymbol{W}_{n_c}\).

To extract the two-loop ultra-collinear function from known results, we exploit the mode separation in the boosted limit. In this regime, the massive TMD soft function constructed from Wilson lines along the heavy-quark velocities \(v_c\) and \(v_d\) refactorizes into a massless TMD soft function and two ultra-collinear functions,
\begin{equation}\label{soft refac}
    \boldsymbol{S}_{cd}^m(b_x, m_t/Q, y^*,\mu ) = \boldsymbol{S}_{cd}(b_x,y^*, \mu, \nu ) S_c^{\mathrm{uc}}(b_x, m_t, \mu, \nu )S_d^{\mathrm{uc}}(b_x, m_t, \mu, \nu ).
\end{equation}
The soft function $\boldsymbol{S}_{cd}^m$ denotes the vacuum expectation values of the two soft Wilson lines along the $v_c,v_d$ directions, which is defined as
\begin{equation}
    \boldsymbol{S}_{cd}^m(b_x, m_t/Q, y^*,\mu ) = \left\langle0\right|\overline{\mathbf{T}}[\boldsymbol{S}_{v_d}^\dagger\left(b^\mu\right)\boldsymbol{S}_{v_c}\left(b^\mu\right)]\mathbf{T}[\boldsymbol{S}_{v_c}^{\dagger}(0)\boldsymbol{S}_{v_d}(0)]|0\rangle.
\end{equation}
For the process of $Q\bar{Q}V$ production at lepton colliders, the soft function $\boldsymbol{S}_{cd}^m$ can be directly obtained from the fully differential soft function computed to the two loops in Ref.~\cite{Liu:2024hfa}. On the right-hand side of Eq.~\eqref{soft refac}, $\boldsymbol{S}_{cd}$ represents the standard massless TMD soft function
\begin{equation}
    \boldsymbol{S}_{cd}(b_x, y^*,\mu,\nu ) = \left\langle0\right|\overline{\mathbf{T}}[\boldsymbol{S}_{n_d}^\dagger\left(b^\mu\right)\boldsymbol{S}_{n_c}\left(b^\mu\right)]\mathbf{T}[\boldsymbol{S}_{n_c}^{\dagger}(0)\boldsymbol{S}_{n_d}(0)]|0\rangle.
\end{equation}
By expanding $\boldsymbol{S}_{cd}^m$ in the boosted limit and matching Eq.~\eqref{soft refac} order-by-order in $\alpha_s$, we extract the renormalized two-loop ultra-collinear function in the form
\begin{equation}
        S_c^{\mathrm{uc},(2)}(b_x, m_t, \mu, \nu ) 
        = \left(\frac{\alpha_s^{(n_l)}}{4\pi}\right)^2 
        C_F \left(C_F\mathcal{K}_{C_F}+C_A\mathcal{K}_{C_A}+n_lT_F\mathcal{K}_{n_lT_F}+c_2\right),
\end{equation}
where the coefficients are
\begin{equation}
    \begin{aligned}
        \mathcal{K}_{C_F}=&\left(\frac{1}{2}L_b^4
        + \big(2 + 2L_{cs}\big)L_b^3
        + \Big(2+\frac{\pi^2}{6} + 4L_{cs} + 2L_{cs}^2\Big)L_b^2 + \frac{\pi^2}{3}\big(1+L_{cs}\big)L_b\right),\\
        \mathcal{K}_{C_A}=&\left(\frac{22}{9} L_b^3 
+ \left(\frac{11}{3}L_{cs}+\frac{100}{9} - \frac{\pi^2}{3}\right) L_b^2\right.\\
&\left.+\left( \left(\frac{134}{9} - \frac{2}{3}\pi^2\right) L_{cs}+\frac{294}{27} + \frac{5\pi^2}{9} + 4\zeta_3\right) L_b+\left(\frac{404}{27}-14\zeta_3\right)L_{cs}
\right),\\
        \mathcal{K}_{n_lT_F}=&\left(-\frac{8}{9} L_b^3
-\left(\frac{32}{9} + \frac{4}{3} L_{cs}\right) L_b^2
-\left(\frac{40}{9} + \frac{4\pi^2}{9} + \frac{40}{9} L_{cs}\right) L_b-\frac{112}{27}L_{cs}\right).
    \end{aligned}
\end{equation}
Here, $L_{cs}=\ln(\nu^2 m_t^2/(\mu^2Q^2))$, and $c_2=137.1935-9.3216n_lT_F$. We reconstructed the analytic coefficients of all logarithms, while the constant terms were determined numerically. We verified the consistency of Eq.~\eqref{soft refac} by checking the \(\mu\)- and \(\nu\)-independence, confirming its validity for both lepton and hadron colliders. For hadronic applications, the corresponding result is obtained by the substitution \(L_{cs}\to \ln(\nu^2/\zeta_{c(d)})\).

The renormalized ultra-collinear function satisfies the RG and RRG equations
\begin{equation}
\begin{aligned}
\frac{\mathrm{d}}{\mathrm{d} \ln \mu}S_c^{\mathrm{uc}}(b_x, m_t, \mu, \nu )&= \Gamma_c^{\mathrm{uc}}\left(\mu\right)S_c^{\mathrm{uc}}(b_x, m_t, \mu, \nu),\\
\frac{\mathrm{d}}{\mathrm{d} \ln \nu}S_c^{\mathrm{uc}}(b_x, m_t, \mu, \nu )&= \Gamma^{\mathrm{uc}}_{\nu}\left(\mu\right)S_c^{\mathrm{uc}}(b_x, m_t, \mu, \nu),
\end{aligned}
\end{equation}
with the anomalous dimensions
\begin{equation}
\begin{aligned}\label{AD:Uc}
   \Gamma_c^{\mathrm{uc}}\left(\mu\right)&=C_F \gamma_{\text {cusp }}^{(n_l)}\left(\mu\right) \ln \frac{m_t^2\nu^2}{\omega_{c}^2\mu^2}+\gamma_{\mathrm{uc}}^{(n_l)},\\
    \Gamma^{\mathrm{uc}}_\nu\left(\mu\right)&=-C_F\left[ 2A^{(n_l)}_\Gamma(b_0/b_x,\mu) - \gamma_{\mathrm{uc},\nu}^{(n_l)}(b_0/b_x) \right].
   \end{aligned}
\end{equation}
The rapidity anomalous dimension satisfies the relation $\gamma^{(n_l)}_{\mathrm{uc},\nu}= -\gamma^{(n_l)}_r$.

\subsection{Resummation formula} \label{secResummation}

With the anomalous dimensions presented for all the ingredients, we demonstrate that our factorization formula, given in Eq.~\eqref{cross section}, satisfies the consistency relations required by RG invariance. To facilitate the final result, we introduce the subtracted soft function $\tilde{\boldsymbol{S}}_{ab \to cd}(b_x, y^*, \mu, \nu) \equiv \boldsymbol{S}_{ab \to cd}(b_x, y^*, \mu, \nu) / S_{ab}(b_x, \mu, \nu)$. By combining the solutions of the evolution equations for each component, we obtain the resummed expression for the azimuthal angular distribution as
\begin{align}
\label{resformu}
& \frac{\mathrm{d}^4 \sigma }{\mathrm{~d} y_c \mathrm{~d} y_d \mathrm{~d} p_T^2 \mathrm{~d} \delta \phi}=\sum_{a b} \frac{x_a x_b p_T}{16 \pi \hat{s}^2} \int_0^{\infty} \frac{2 \mathrm{~d} b_x}{\pi} \cos \left(b_x p_T \delta \phi\right)   \nonumber\\
&\times \tilde f_{a / p}\left(x_a, b_*, \mu_{b_*}, \zeta_{a, i}\right)  \tilde f_{b/ p}\left(x_b, b_*, \mu_{b_*}, \zeta_{b, i}\right)\left(\sqrt{\frac{\zeta_{a, f}}{\zeta_{a, i}}}\right)^{\kappa_a(b_*, \mu_{b_*})}\left(\sqrt{\frac{\zeta_{b, f}}{\zeta_{b, i}}}\right)^{\kappa_b(b_*, \mu_{b_*})}\nonumber\\
& \times \exp \left\{-\int_{\mu_{j}}^{\mu_h} \frac{\mathrm{d} \mu}{\mu}\left[\gamma_{\text {cusp }}^{(n_f)}\left(\mu\right) C_H \ln \frac{\hat{s}}{\mu^2}+2 \gamma_H^{(n_f)}\left(\mu\right)\right]\right\} \nonumber\\
& \times \exp \left\{-\int_{\mu_{b_*}}^{\mu_j} \frac{\mathrm{d} \mu}{\mu}\left[\gamma_{\text {cusp }}^{(n_l)}\left(\mu\right) C_H \ln \frac{\hat{s}}{\mu^2}+2 \gamma_H^{(n_l)}\left(\mu\right)\right]\right\} \nonumber\\
& \times \sum_{K K^{\prime}} \exp \left[-\int_{\mu_{j}}^{\mu_h} \frac{\mathrm{d} \mu}{\mu}\gamma_{\text {cusp }}^{(n_f)}\left(\mu\right)\left(\lambda_K+\lambda_{K^{\prime}}^*\right)\right]\nonumber \\
&\times  \exp \left[-\int_{\mu_{b_*}}^{\mu_j} \frac{\mathrm{d} \mu}{\mu}\gamma_{\text {cusp }}^{(n_l)}\left(\mu\right)\left(\lambda_K+\lambda_{K^{\prime}}^*\right)\right] H_{K K^{\prime}}\left(M, y^*, \mu_h\right) \tilde S_{K^{\prime} K}\left(b_*,y^*, \mu_{b_*},\nu_s\right)  \nonumber\\
& \times \exp \left[-\int_{\mu_{b_*}}^{\mu_j} \frac{\mathrm{d} \mu}{\mu} \left(\Gamma^{\mJ_c}\left(\mu\right)+\Gamma^{\mJ_d}\left(\mu\right)\right)\right] \mJ_c(m_t,\mu_j,\nu_{m,c})\mJ_d(m_t,\mu_j,\nu_{m,d}) \nonumber\\
&\times \exp \left[-\int_{\mu_{b_*}}^{\mu_j} \frac{\mathrm{d} \mu}{\mu} \Gamma^{\mS_m}\left(\mu\right)\right]\mS_m(m_t,\mu_j,\nu_m) \left(\frac{\nu_m^2}{\nu_{m,c}\,\nu_{m,d}}\right)^{\Gamma_\nu^{\mJ_m}(\mu_{b_*})}\nonumber\\
& \times S_c^{\mathrm{uc}}(b_*, m_t, \mu_{b_*}, \nu_c )S_d^{\mathrm{uc}}(b_*, m_t, \mu_{b_*}, \nu_d ) \left(\frac{\nu_s^2}{\nu_c\,\nu_d}\right)^{\Gamma^{\mathrm{uc}}_\nu(\mu_{b_*})}\nonumber\\
& \times \exp \left[-S_{\mathrm{NP}}^a\left(b_x, Q_0, \sqrt{\hat{s}}\right)-S_{\mathrm{NP}}^b\left(b_x, Q_0, \sqrt{\hat{s}}\right)\right].
\end{align}
Here, the differential over transverse momentum $q_x$ has been converted to the azimuthal decorrelation $\delta\phi$, and $\mu_h$, $\mu_j$, and $\mu_{b_*}$ correspond to the hard, collinear, and soft scales, respectively. A key feature of Eq.~\eqref{resformu} is the implementation of a segmented evolution scheme~\cite{Hoang_2015}, which properly accounts for the changing number of active flavors across the heavy-quark threshold. Specifically, the evolution from the hard scale $\mu_h$ down to $\mu_j$ is performed in the $n_f=6$ flavor scheme, while the subsequent evolution from $\mu_j$ to $\mu_{b_*}$ utilizes the $n_l=5$ flavor scheme. To resolve the color mixing effects in the evolution, we diagonalize the anomalous dimension matrix $\boldsymbol{M}_{ab\to cd}$. The quantities $\lambda_K$ represent the eigenvalues of the matrices $\boldsymbol{M}_{1,2}$, while $H_{K K^{\prime}}$ and $\tilde S_{K^{\prime} K}$ denote the components of the hard and soft functions in this diagonalized basis.

The remaining factors describe the evolution of the specific functions. The TMDPDFs evolve via the CSS formalism. The massive jet and massive soft functions evolve from $\mu_j$ to $\mu_{b_*}$ using the standard RG and RRG equations in the $n_l$-flavor scheme, where $\nu_{m,c(d)}$ and $\nu_m$ denote their respective rapidity scales. Similarly, the ultra-collinear functions are evolved to the factorization scales, with $\nu_{c(d)}$ and $\nu_s$ representing the typical rapidity scales for the ultra-collinear and soft sectors. Finally, the non-perturbative Sudakov factors $S_{\rm NP}$ parameterize the intrinsic motion of the bound partons at the scale $Q_0$~\cite{Sun:2014dqm, Kang:2015msa, Echevarria:2020hpy, Alrashed:2021csd}. The $b_*$-prescription is adopted to regularize the Landau pole singularity, as detailed in Sec.~\ref{sec:num}.

\section{Numerical results}
\label{sec:num}

In this section, we present numerical predictions derived from the resummation formula in Eq.~\eqref{resformu}.  We apply this formalism to boosted top-quark pair production to describe the azimuthal decorrelation of the final-state particles.  Our framework systematically resums large logarithms arising from both the small azimuthal angle limit and boosted heavy-quark limit, achieving $\mathrm{NNLL}^\prime$ accuracy.

Following standard LHC analyses \cite{Kaplan:2008ie, CMS:2014fya, ATLAS:2017jiz}, we consider a center-of-mass energy of $\sqrt{S}=13\,\text{TeV}$ and restrict the final-state top quarks to the central rapidity region, $|y_{c,d}| < 2$, with a transverse momentum cut of $p_T > 400\,\text{GeV}$. To focus on the regime dominated by all-order resummation effects, we restrict our analysis to the azimuthal range $0 < \delta\phi < 0.4$.

To regularize the Landau pole in the non-perturbative regime, we adopt the standard $b_*$-prescription~\cite{Collins:1984kg}, defining the regulated impact parameter and associated scale as 
\begin{align}
b_* &\equiv \frac{|b_x|}{\sqrt{1+b_x^2/b_{\rm max}^2}}, \quad \mu_{b_*} = \frac{2 e^{-\gamma_E}}{b_*}.
\end{align}
Following Refs.~\cite{Sun:2014dqm, Kang:2015msa, Echevarria:2020hpy,  Alrashed:2021csd}, the non-perturbative Sudakov factors for the TMDPDFs in Eq.~\eqref{resformu} are parameterized as
\begin{align}
S_{\rm{NP}}^{i} = g_1\, b_x^2 + \frac{g_2}{2}\frac{C_i}{C_F}\ln\frac{Q}{Q_0}\ln\frac{|b_x|}{b_*},
\end{align}
where $i \in \{a,b\}$. We utilize the fitted parameters $g_1 = 0.106\,\text{GeV}^2$, $g_2 = 0.84$, and $Q_0^2 = 2.4\,\text{GeV}^2$. The canonical scales entering the resummation formula are chosen as follows
\begin{align}\label{scales}
\mu_h = 2 p_T, \quad \mu_j = m_t, \quad &\nu_s = \mu_{b_*}, \quad \nu_i = \omega_{i} \mu_{b_*}/m_t, \quad \nu_{m,i} = \omega_i, \quad \nu_m = m_t,\notag\\
\zeta_{a,f}& = \zeta_{b,f} = \hat{s}, \quad \zeta_{a,i} = \zeta_{b,i} = b_0^2/b_x^2,
\end{align}
with the top-quark mass set to $m_t=172.5\,\text{GeV}$.

\begin{figure}[t]
    \centering
    \includegraphics[width=15cm]{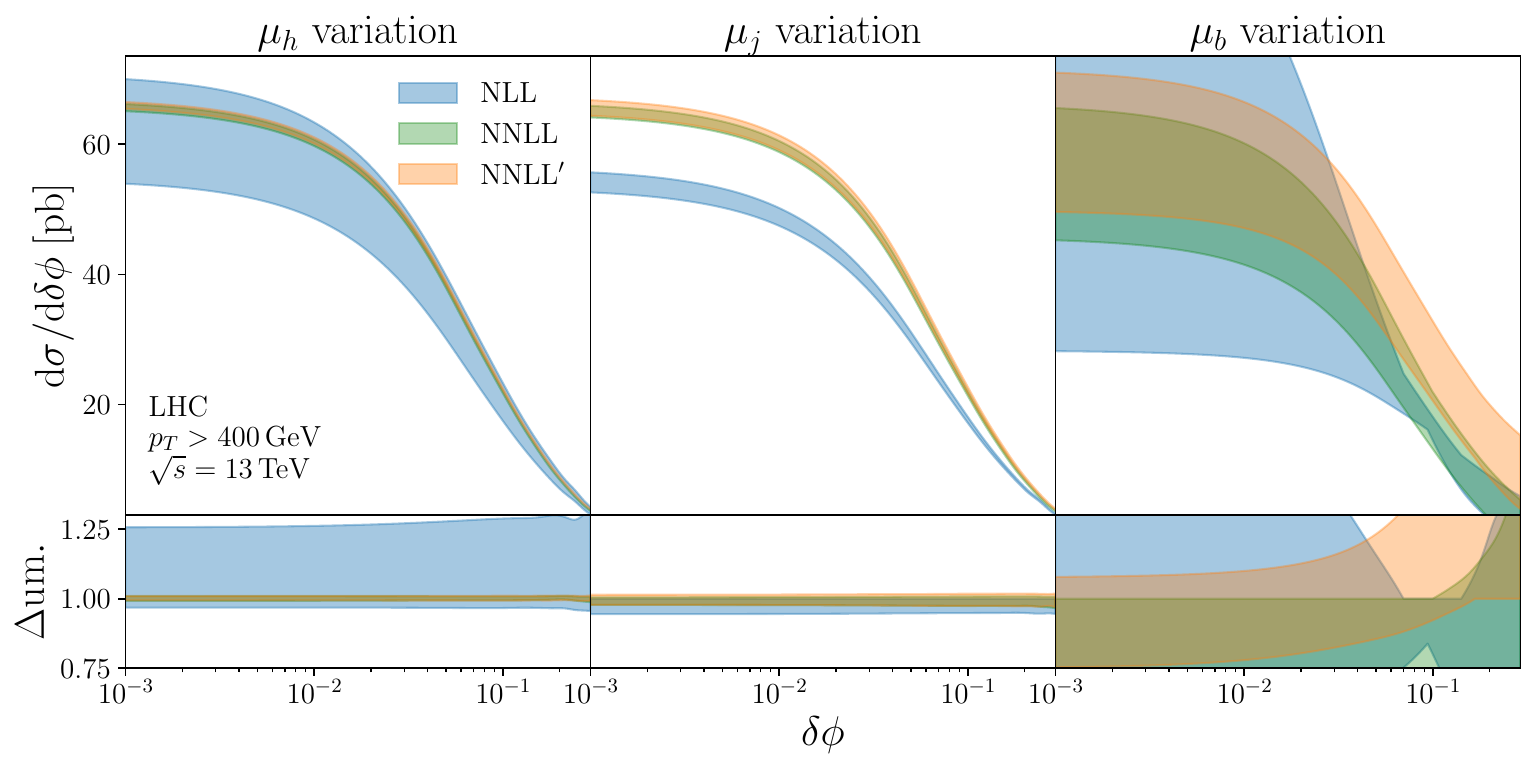}
    \caption{Azimuthal $\delta\phi$ distribution in $pp$ collisions at NLL (blue), NNLL (green) and NNLL$^\prime$ (orange) accuracy. The uncertainty estimates are shown for variations in the renormalization scales $\mu_h$ and $\mu_j$ and $\mu_{b_{*}}$, where each scale is varied up and down by a factor of two.} 
    \label{fig:res_results}
\end{figure}

In Fig.~\ref{fig:res_results}, we present the resummed azimuthal angular distribution at NLL (blue), NNLL (green), and $\mathrm{NNLL}^\prime$ (orange) accuracy. Theoretical uncertainties are estimated via independent variations of the hard scale $\mu_h$, the collinear scale $\mu_j$, and the soft scale $\mu_{b_*}$ by a factor of two around their canonical values. The uncertainty bands exhibit a substantial reduction from NLL to NNLL. However, in moving from NNLL to $\mathrm{NNLL}^\prime$, the width of the uncertainty bands remains comparable. This behavior is consistent with the definition of ``prime'' accuracy: while $\mathrm{NNLL}^\prime$ incorporates fixed-order matching coefficients at one higher loop order ($\mathcal{O}(\alpha_s^2)$), it utilizes the same evolution kernels as NNLL. Consequently, the central values are refined by these higher-order boundary conditions, but the scale dependence does not decrease significantly.

Additionally, we observe that the reduction in scale uncertainty is less pronounced for variations of $\mu_{b_*}$ and $\mu_j$ compared to $\mu_h$. This sensitivity arises because $\mu_{b_*}$ and $\mu_j$ lie closer to the non-perturbative regime. In the back-to-back limit $(\delta\phi \rightarrow 0)$, the resummation formula effectively resolves the singularities intrinsic to fixed-order perturbation theory, yielding a finite, physical prediction. As the azimuthal separation $\delta\phi$ increases, power-suppressed corrections of $\mathcal{O}(\delta\phi)$ become significant. These terms lead to a degradation in accuracy outside the back-to-back region. While matching to fixed-order calculations would extend the validity of the prediction to the full kinematic range, the present study focuses on the resummation-dominated regime where these large logarithms are paramount.

\section{Conclusions}\label{sec:Conclusions}
In this work, we have established a TMD factorization and resummation framework for boosted top quark pair production in the kinematic regime $p_T \gg m_t \gg |q_x|$ at the LHC. By employing a two-step EFT matching procedure, transitioning from QCD to $\mathrm{SCET}\,+\,\mathrm{HQET}$ and subsequently to $\mathrm{SCET}\,+\,\mathrm{bHQET}$, we have systematically resummed the large logarithms associated with both the top quark mass and the azimuthal decorrelation. A central contribution of this study is the first extraction of the two-loop ultra-collinear function from the fully differential massive soft function.

With all known perturbative ingredients included, we have achieved $\mathrm{NNLL}^\prime$ precision for the azimuthal decorrelation distribution in boosted $t\bar{t}$ production. This represents the highest level of theoretical control currently available for this observable. Our numerical results provide robust predictions with quantified theoretical uncertainties, establishing a benchmark for future phenomenological studies. We emphasize that the results presented here focus on the resummation of singular logarithms in the peak region. The matching of these resummed predictions with fixed-order calculations, essential for a complete description of the far-tail region, is beyond the scope of the current analysis and will be addressed in future work.

Looking ahead, our framework lays a clear pathway toward $\mathrm{N}^3\mathrm{LL}$ accuracy. Most of the requisite anomalous dimensions are already known, with the primary missing ingredients being the three-loop rapidity anomalous dimensions for the massive jet and soft functions. Beyond immediate phenomenological applications for the LHC, our results demonstrate the efficacy of combining SCET and bHQET to tackle multiscale problems in heavy quark physics. The framework is directly applicable to boosted bottom quark pair production, offering timely theoretical predictions for heavy flavor experiments at RHIC. Furthermore, it can be readily generalized to proton nucleus collisions by substituting the proton TMDPDFs with their nuclear modified counterparts. These capabilities, combined with the rigorous treatment of mass effects at $\mathrm{NNLL}^\prime$, open new directions for precision QCD studies in boosted kinematics and provide valuable theoretical input for ongoing and forthcoming experimental analyses.

\acknowledgments
We thank Ze Long Liu for helpful discussions. This work is supported by the National Science Foundations of China under Grant No.~12275052, No.~12147101, No.~12547102, and the Innovation Program for Quantum Science and Technology under grant No. 2024ZD0300101. 

\appendix

\section{Anomalous dimension}\label{app:a-dim}
All the anomalous dimensions  have the following perturbative expansion in the strong coupling constant
\begin{equation}
\gamma_i(\alpha_s) = \sum_{n=0}^{\infty} \gamma^i_{n} \left( \frac{\alpha_s}{4\pi} \right)^{n+1}, \quad \text{with} \quad i = \text{cusp}, q,g, \mJ_m, \text{uc},r , 
\end{equation}
and at NNLL$^\prime$ the cusp anomalous dimension is given by \cite{Moch:2004pa}
\begin{align}
\gamma_0^{\text{cusp}} &= 4, \notag \\
\gamma_1^{\text{cusp}} &= 4 \left[ C_A \left( \frac{67}{9} - \frac{\pi^2}{3} \right) - \frac{20}{9} T_F n_f \right], \notag \\
\gamma_2^{\text{cusp}} &= 4 \left[ C_A^2 \left( \frac{245}{6} - \frac{134 \pi^2}{27} + \frac{11 \pi^4}{45} + \frac{22 \zeta_3}{3} \right) + C_A T_F n_f \left( -\frac{418}{27} + \frac{40 \pi^2}{27} - \frac{56 \zeta_3}{3} \right) \right. \notag \\
& \quad \left. + C_F T_F n_f \left( -\frac{55}{3} + 16 \zeta_3 \right) - \frac{16}{27} T_F^2 n_f^2 \right].
\end{align}
At the same accuracy, the non-cusp quark and gluon anomalous dimensions read in \cite{Moch:2005id, Moch:2005tm, Idilbi:2005ni, Idilbi:2006dg, Becher:2006mr}
\begin{align}
    &\gamma_0^q=-3C_F,\notag \\
    &\gamma_1^q=C_F^2\left(-\frac{3}{2}+2\pi^2-24\zeta_3\right)+C_FC_A\left(-\frac{961}{54}-\frac{11\pi^2}{6}+26\zeta_3\right)+C_FT_Fn_f\left(\frac{130}{27}+\frac{2\pi^2}{3}\right), \notag \\
    &\gamma_{0}^{g}=-\frac{11}{3}C_{A}+\frac{4}{3}T_{F}n_{f}, \notag \\
    &\gamma_{1}^{g}=C_A^2\left(-\frac{692}{27}+\frac{11\pi^2}{18}+2\zeta_3\right)+C_A T_F n_f\left(\frac{256}{27}-\frac{2\pi^2}{9}\right)+4C_FT_Fn_f.
\end{align}
The coefficients in the expansion of the anomalous dimension of $\mJ_m$ are given by~\cite{Neubert:2007je,Hoang_2015}
\begin{align}
\notag &\gamma_0^{\mJ_m} = 2\cf, \\
\notag &\gamma_1^{\mJ_m} = C_F^2 \left( 3 - 4\pi^2 + 48 \zeta_3 \right) + C_F C_A \left( \frac{373}{27} + 5\pi^2 - 60 \zeta_3 \right) - C_F T_F n_l \left( \frac{20}{27} + \frac{4\pi^2}{3} \right).
\end{align}
The anomalous dimensions of ultra-collinear function are
\begin{align}
    \notag \gamma_0^{\mathrm{uc}}&=4\cf,\\
    \gamma_1^{\mathrm{uc}}&=-\cf T_F n_l\left(\frac{16}{27}+\frac{4\pi^2}{9}\right)-\cf\ca\left(\frac{220}{27}+\frac{\pi^2}{9}-36\zeta_3\right);\notag \\
    \gamma_{\nu,0}^{\mathrm{uc}}&=-\gamma_0^r=0, \notag \\
    \gamma_{\nu,1}^{\mathrm{uc}}&=-\gamma_1^r=- C_{A}\left(28\zeta_{3} - \frac{808}{27}\right) - \frac{224T_Fn_{f}}{27}.
\end{align}

\bibliographystyle{jhep}
\bibliography{biblio.bib}

\end{document}